# Biocompatibility of Nanomaterials in Medical Applications

**Marvellous O. Eyube[1,2*], Courage Enuesueke[1,2], Marvellous Alimikhena[1,2]**

[1]*Department of Chemistry, Faculty of Physical Science, University of Benin, Nigeria*

[2]*S.T.E.L.L.A.R. Labs (Science and Technological Enhanced Laboratory for Advance Learning and Research), Benin City, Nigeria*

**Corresponding Author Email:** marvellous.eyube@physci.uniben.edu

Corresponding Author ORCID: https://orcid.org/0009-0003-7951-1830

**Abstract:**

Biocompatibility is a critical factor in the application of nanomaterials in medical fields, as these materials must interact safely and effectively with biological systems to be viable for therapeutic and diagnostic use. This article investigates the biocompatibility of nanomaterials, focusing on their interactions with biological cells, tissues, and the immune system. Key properties such as surface chemistry, size, shape, and material composition are examined, as they significantly influence the biological response. The article explores the role of nanomaterials in medical applications, including drug delivery, diagnostic imaging, and tissue engineering, while discussing the challenges involved in enhancing their biocompatibility. A case study on the CaO-CaP binary system is presented, showcasing the use of calcium oxide (CaO) and calcium phosphate (CaP) nanoparticles in bone tissue engineering. This system is widely investigated for its ability to mimic the mineral content of bone and promote osteogenesis, highlighting both its therapeutic potential and challenges in ensuring safe biocompatibility in clinical settings. The article concludes by reviewing strategies to optimize the biocompatibility of nanomaterials and discussing future directions for research in advancing their applications in medical treatments.

# 1. Introduction and the Medical Demand for Nanomaterials

Modern medicine is witnessing a paradigm shift, shaped by the rise of precision medicine, implantable technologies, and patient-specific treatment regimens. These emerging approaches demand materials that can perform reliably within complex biological systems while enabling fine-tuned control over therapeutic or diagnostic outcomes. However, conventional biomaterials often fall short of these requirements. Their limited biological responsiveness, low adaptability, and potential to trigger immune reactions have created a critical gap in achieving next-generation medical solutions [1].

Nanomaterials have emerged as transformative candidates capable of addressing these challenges due to their unique structural and functional properties. Their nanoscale dimensions allow interaction with biomolecules, cells, and tissues at fundamental biological levels. Features such as large surface-area-to-volume ratio, tunable surface chemistry, and responsiveness to external stimuli give them a degree of biofunctionality that conventional materials lack [2]. Furthermore, their versatility allows them to be integrated into various platforms—from injectable drug delivery systems to implant coatings—while scalability in production offers potential for widespread clinical application [3].

These advantages make nanomaterials particularly attractive for real-world medical demands. In cancer therapy, for example, liposomal formulations such as Doxil® have revolutionized chemotherapy by delivering doxorubicin directly to tumor sites, reducing systemic toxicity and improving therapeutic efficacy. This milestone—recognized as the first FDA-approved nanodrug—demonstrates how rational nanodesign can overcome longstanding limitations in pharmacokinetics and safety [4]. In diagnostic imaging, ferumoxytol, an iron oxide nanoparticle, has been successfully used off-label as an MRI contrast agent in various clinical settings, enhancing vascular imaging in patients for whom conventional gadolinium-based agents are unsuitable [5]. These examples underscore the ability of nanotechnology to enhance performance and expand the clinical toolbox across disciplines.

In orthopedic and dental applications, nanostructured coatings and scaffolds—such as those made from calcium phosphate or calcium oxide—promote bone regeneration and tissue integration due to their osteoconductive nature [6]. More specifically, nano-hydroxyapatite (nHAp) scaffolds have shown great promise in clinical and preclinical bone regeneration efforts, offering high surface reactivity, biomineral mimicry, and superior compatibility with osteoblasts [7]. These scaffolds not only provide structural support but also serve as bioactive matrices that modulate cell behavior and promote osteogenesis.

Another critical area of application is tissue engineering, where nanomaterials mimic the extracellular matrix and support cellular activities necessary for tissue regeneration. These scaffolds, with high surface area and controllable porosity, facilitate the attachment of cells and the localized delivery of bioactive agents, enabling the repair of damaged or diseased tissues [8]. Across all these fields, nanomaterials are not only enhancing current medical practices but also enabling technologies that were previously unachievable with traditional systems.

Despite these promising developments, the clinical use of nanomaterials hinges on their ability to safely interact with biological environments. Their high reactivity, while beneficial for functionality, introduces risks of cytotoxicity, inflammation, or immune system activation. As such, biocompatibility has emerged as a core requirement for successful application in medicine. Defined as a material's ability to perform its intended role without provoking adverse biological responses, biocompatibility ensures that nanomaterials are both effective and safe for clinical use [1,2].

The next section explores this concept in depth—examining the criteria for assessing biocompatibility, the mechanisms by which nanomaterials interact with biological systems, and the strategies employed to mitigate risks. As the foundation for all subsequent medical integration, biocompatibility forms the critical bridge between nanomaterial innovation and real-world patient outcomes.

## 2. The Critical Significance of Biocompatibility in Nanomedicine

### 2.1. The Role of Biocompatibility in Clinical Success

Biocompatibility is central to the successful application of nanomaterials in medicine. In clinical settings, it ensures that materials interact with biological tissues without triggering adverse reactions such as toxicity, inflammation, or immune rejection. Nanomaterials that are not biocompatible may provoke acute or chronic responses, potentially leading to the failure of a therapy or device. As medicine increasingly relies on materials that operate at the molecular and cellular scale, biocompatibility becomes not just desirable, but essential for clinical efficacy and patient safety [9].

Instances of therapeutic failure due to immune activation, accelerated clearance, or unforeseen toxicity highlight the need to prioritize biocompatibility early in nanomaterial design. For example, nanoparticles that perform well under laboratory conditions may behave unpredictably in vivo if their surface is not tailored to evade immune surveillance or protein fouling. These failures underscore the importance of understanding how nanomaterials interact with biological systems in order to reduce the likelihood of clinical setbacks [10].

## 2.2. Comparison with Conventional Biomaterials

Nanomaterials differ significantly from traditional biomaterials, particularly in their interaction with biological systems. Unlike bulk materials that are often inert and mechanically focused, nanoscale materials exhibit high surface energy, increased reactivity, and tunable physicochemical properties, all of which impact their biocompatibility. These properties enable unprecedented medical functionalities but also introduce challenges such as protein adsorption, immune activation, and cytotoxicity if not properly controlled [11].

In contrast to conventional implants made of metals or polymers, nanomaterials are often designed for more dynamic roles, such as drug delivery or real-time biosensing. Their smaller size enhances tissue penetration and cellular uptake but also increases the surface area available for interactions with proteins, lipids, and immune cells. For instance, nanoparticles under 100 nm are efficiently taken up by cells but may accumulate in tissues or organs, leading to long-term toxicity if they are not biodegradable [12]. Additionally, particle shape influences behavior—spherical particles are generally taken up more readily than rod-like ones, which may persist longer and interact differently with immune pathways [13].

## 2.3. Regulatory Emphasis on Biocompatibility

Regulatory bodies such as the U.S. Food and Drug Administration (FDA), the European Medicines Agency (EMA), and the International Organization for Standardization (ISO) have emphasized biocompatibility as a prerequisite for nanomaterial approval. Beyond proving therapeutic benefit, developers must provide robust data on toxicity, immunogenicity, biodegradation, and clearance profiles. These factors are particularly scrutinized for nanomaterials due to their complexity and the evolving understanding of their interactions with biological systems [14].

Designing for regulatory compliance involves thoughtful material selection and surface engineering to enhance safety. Organic nanomaterials, for example, often break down into biocompatible byproducts, making them more favorable for long-term applications. In contrast, inorganic systems may require surface modifications—such as coating or encapsulation—to meet safety thresholds and prevent accumulation or chronic toxicity [15,16].

## 2.4. Foundation for Engineering Nanomedicines

Biocompatibility is not an afterthought—it must be embedded into the nanomaterial's design from the start. Surface chemistry, size, shape, charge, and composition all govern how a nanomaterial interacts with cells and tissues. One common method to improve compatibility is surface functionalization, such as PEGylation, which extends circulation time and minimizes recognition by immune cells [10,11]. Surface hydrophilicity is also crucial, as hydrophilic coatings reduce non-specific protein adsorption and enhance cellular interactions [11].

Surface charge, in particular, plays a dual role. Positively charged nanomaterials can enhance cellular uptake due to electrostatic interactions with negatively charged membranes but may simultaneously increase cytotoxicity and inflammation [17]. Negatively charged or neutral surfaces, though less aggressive in uptake, often present reduced immunogenicity. Achieving a balanced surface charge is thus critical in optimizing therapeutic performance while minimizing risks [17].

Another essential parameter is surface energy, which influences how nanomaterials interact with proteins and form a biological identity—or "protein corona"—upon entering the bloodstream. This corona can alter distribution, cellular uptake, and immune responses. Modulating surface energy through chemical design can either suppress or guide these interactions in favor of the intended application [18].

Material composition further influences biocompatibility. Organic nanomaterials, such as liposomes and biodegradable polymers, are typically well-tolerated and degrade into safe byproducts, making them suitable for sustained or repeated administration [14]. On the other hand, inorganic materials like silica, iron oxide, or gold offer mechanical or imaging advantages but may require surface modifications to mitigate potential toxicity or long-term accumulation [15]. These considerations have become particularly important in antiviral nanomedicine, where tailored surface properties—such as charge, hydrophilicity, and coating strategies—are now central to ensuring both therapeutic efficacy and immunological safety in viral prevention and treatment platforms [19].

## 3. Methodologies for Biocompatibility Assessment

Evaluating the biocompatibility of nanomaterials requires a multidisciplinary approach involving experimental assays, animal studies, computational modeling, and regulatory frameworks. These methodologies collectively assess how nanomaterials interact with biological systems, offering a comprehensive safety and efficacy profile tailored to their medical applications. Given the complexity of nano–bio interactions, integrating these varied approaches helps researchers mitigate risks while advancing clinical translation.

### 3.1. In Vitro Methods

In vitro techniques represent the initial screening tools for assessing the biocompatibility of nanomaterials. These laboratory-based assays evaluate the cellular responses to nanoparticles under controlled conditions, eliminating the complexity of whole-organism interactions. Common methods include cell viability assays such as MTT and resazurin reduction, which quantify metabolic activity as an indicator of cytotoxicity. Additionally, membrane integrity tests, oxidative stress assessments, and apoptosis detection are employed to reveal subcellular impacts of nanomaterials [20].

Real-world applications have demonstrated the efficacy of in vitro tools in nanomedicine research. For instance, Siller *et al.* (2019) developed a real-time live-cell imaging system to monitor cytotoxicity and cellular morphology in response to 3D-printed biomaterials—enabling high-throughput and time-resolved assessment of biocompatibility [21]. Another case involved the evaluation of zinc oxide nanoparticles synthesized from *Artemisia annua*, where MTT assays confirmed enhanced osteogenic differentiation in human osteoblast-like MG-63 cells [22].

Advanced in vitro systems, including 3D cell cultures and co-culture models, better mimic the tissue microenvironment compared to traditional 2D monolayers. These systems allow the observation of nanoparticle-induced changes in cell proliferation, differentiation, and inflammatory signaling. Although in vitro studies are cost-effective and high-throughput, they lack the complexity of physiological systems—necessitating follow-up in vivo studies to confirm findings [20].

### 3.2. In Vivo Methods

In vivo approaches evaluate biocompatibility within living organisms, providing vital insights into systemic distribution, metabolism, clearance, and potential toxicity. Rodent models, such as mice and rats, are commonly used to assess both acute and chronic biological responses. These studies help identify critical endpoints including immune activation, hematological changes, and organ-specific toxicities [23].

An exemplary application is the use of CaO–CaP nanocomposites in Wistar rats, where scaffold implantation in bone defects led to observable tissue regeneration and immune modulation. Moreover, histopathological examination—central to in vivo assessment—enabled detection of fibrosis, necrosis, and inflammation across major organ systems. This approach is further supported by Kyriakides *et al.* (2021)**,** who provided comprehensive insights into the immunological outcomes of various nanomaterials in vivo, including cytokine induction, complement activation, and tissue compatibility [24].

Despite their importance, in vivo methods raise ethical concerns and are limited by species-specific differences that may not fully predict human responses. Consequently, emerging platforms like organ-on-a-chip and ex vivo perfusion systems are gaining traction as ethical and functional alternatives [23].

### 3.3. Computational Models

Computational models serve as predictive tools for anticipating nanomaterial interactions with biological systems. Techniques such as molecular dynamics (MD) simulations and quantitative structure–activity relationship (QSAR) modeling allow researchers to estimate nanoparticle behavior based on their physicochemical properties. MD simulations help model nanoscale interactions, such as membrane disruption or protein binding, offering atomic-level resolution of potential toxicity pathways [25].

Recent developments in nano-QSAR, such as those reported by Cao *et al.* (2020), have enabled risk prediction for metal oxide nanoparticles based on computational descriptors like band gap energy, particle charge, and hydration energy—streamlining early-phase toxicology testing [26].

These models are particularly valuable for screening large nanomaterial libraries without extensive biological assays.

QSAR models correlate descriptors like particle size, surface charge, and hydrophobicity with observed biological outcomes. However, their predictive power depends heavily on the quality and diversity of training datasets. Limitations include the lack of standardized descriptors and difficulty simulating complex biological microenvironments. Continued refinement and validation are essential for regulatory acceptance [25,26].

### 3.4. Regulatory Standards

The regulatory evaluation of nanomaterials is guided by frameworks developed by agencies such as the U.S. Food and Drug Administration (FDA), European Medicines Agency (EMA), and international bodies like the International Organization for Standardization (ISO) and the Organisation for Economic Co-operation and Development (OECD). These organizations mandate rigorous assessment of toxicity, pharmacokinetics, and biocompatibility prior to clinical approval [27].

A landmark example of regulatory success is the approval of Doxil®, the first FDA-approved nanodrug, which underwent extensive in vitro and in vivo biocompatibility testing, including sterility, hemolysis, and immunogenicity studies [28]. However, current regulatory standards often lag behind rapid advancements in nanomedicine. Discrepancies in international guidelines and the lack of harmonized testing protocols remain major barriers to global commercialization.

The regulatory landscape continues to evolve, with agencies pushing for validated in vitro models, robust in silico predictions, and reliable biocompatibility thresholds. This push for standardization enhances safety while supporting innovation [27].

## 3.5. Comparative Metrics and Evaluation Criteria

As multiple assessment techniques are employed, establishing standardized metrics is essential for comparing biocompatibility outcomes across platforms. Key evaluation criteria include cytotoxicity thresholds, inflammatory markers (e.g., IL-6, TNF-α), cellular uptake efficiency, biodistribution profiles, and histological scores. To support objective analysis, benchmark values are increasingly being established across studies, enabling meta-analyses and cross-comparative assessments.

Additionally, efforts are underway to create Integrated Testing Strategies (ITS), which combine in vitro, in vivo, and in silico results into unified decision-making frameworks. These strategies reduce redundancy, improve predictive accuracy, and accelerate regulatory approval. The concept is exemplified by projects like Schloemer *et al.* (2023), who incorporated quantum-based modeling and biological validation to assess triplet-triplet annihilation in nano-upconversion materials [29].

This convergence of biological and computational evaluation strengthens the translational potential of nanomaterials. As highlighted by Seoane-Viaño *et al.* (2021), the successful transition of 3D-printed drug-delivery systems into clinics depends not just on innovative design but also on robust biocompatibility evaluation and regulatory navigation [30].

## 4. Scientific Challenges in Biocompatibility

The rapid integration of nanomaterials into medical science has created opportunities for breakthroughs in diagnostics, therapy, and regenerative medicine. However, despite their promising potential, nanomaterials present scientific and translational challenges related to biocompatibility. These challenges must be addressed to ensure clinical safety, long-term stability, and regulatory approval. This section outlines the key hurdles faced in biocompatible nanomedicine, focusing on toxicity, immune interactions, material degradation, and safety governance.

### 4.1. Toxicity and Immune Response

One of the primary challenges in deploying nanomaterials for medical use is managing their toxicological and immunological effects. Nanoparticles, due to their high surface area-to-volume ratio, can interact intensely with biological components such as proteins, membranes, and immune cells. While these interactions can be therapeutically useful, they may also provoke unintended effects such as oxidative stress, pro-inflammatory cytokine release, or complement activation.

Surface properties such as charge, hydrophobicity, and functional groups significantly influence biocompatibility and immune recognition. For example, positively charged particles may enhance cellular uptake but are also more likely to disrupt membrane integrity or induce inflammation. Even particles deemed safe in vitro may exhibit unforeseen toxicity when exposed to the complex environments of living organisms.

Additionally, the formation of a protein corona on the nanoparticle surface can alter its identity and behavior in vivo, potentially leading to misrecognition by the immune system. These complications highlight the need for careful material design and thorough preclinical testing to mitigate immunotoxic effects.

**4.2. Long-Term Stability and Degradation**

Nanomaterials intended for biomedical applications must demonstrate stability under physiological conditions, as well as controlled biodegradation where applicable. Materials that degrade too quickly may release toxic byproducts, while non-degradable materials may accumulate in tissues, leading to long-term cytotoxicity or organ burden.

For example, certain inorganic nanoparticles, though stable and effective for imaging or therapy, may persist in the body without clear metabolic or excretory pathways. This persistence raises concerns about chronic exposure, particularly for repeated or systemic applications.

Additionally, a major concern with non-biodegradable inorganic nanomaterials is their tendency to accumulate in vital organs such as the liver, spleen, and kidneys—organs that are central to clearance and detoxification. Nanoparticles exceeding 50 nm, particularly gold and iron oxide varieties, often become sequestered in the reticuloendothelial system (RES), where they are difficult to excrete due to limited renal clearance mechanisms. This long-term accumulation can lead to chronic inflammation, oxidative stress, or functional impairment in these organs. To address this challenge, several strategies have emerged: PEGylation of surfaces helps reduce opsonization and RES uptake; microencapsulation can control release and shield particles from immune detection; and downsizing nanoparticles below 10 nm enhances renal filtration and excretion. For example, gold nanoparticles larger than 50 nm show significantly reduced clearance and prolonged retention in hepatic and splenic tissues, as reported by Kumar et al. (2022) [15].

These solutions are vital for ensuring long-term biocompatibility, especially for nanomaterials intended for systemic or repeat-dose applications.

Conversely, biodegradable polymers such as polycaprolactone (PCL) and polylactic acid (PLA) offer a more controlled degradation profile. When engineered appropriately, these materials break down into non-toxic byproducts, minimizing long-term risk. However, predicting and standardizing degradation rates across different biological environments remains an ongoing challenge [31].

### 4.3. Regulatory and Safety Concerns

The development of biocompatible nanomaterials faces additional hurdles at the regulatory interface, where evolving safety standards and fragmented global policies can slow innovation. Regulatory bodies like the FDA, EMA, and ISO have established frameworks for toxicity testing and biocompatibility assessment, but many of these standards were originally designed for bulk materials and do not fully account for nano-specific behaviors.

Moreover, safety assessments must now include data on inter-individual variability, long-term biodistribution, and nanomaterial interaction with complex pathophysiologies, particularly for personalized and implantable therapies. These requirements demand new testing protocols, interdisciplinary evaluations, and ongoing collaboration between researchers, industry stakeholders, and regulatory agencies.

To meet these demands, next-generation nanomedicines must be developed with regulatory foresight, ensuring that biocompatibility is not only optimized at the design level but also validated against globally recognized safety benchmarks.

# 5. Strategies for Targeted Improvement

Optimizing nanomaterials for medical use requires deliberate strategies that address biocompatibility at the design stage. These strategies not only minimize undesirable biological interactions but also enhance therapeutic precision, safety, and clinical viability. While many of these approaches have been extensively validated in laboratory settings, their translation into real-world systems remains a critical measure of success. This section outlines four core strategies—surface modification, use of biodegradable platforms, targeted delivery, and hybrid nanostructures—that lay the foundation for more compatible and functional nanomedical technologies. These principles are further exemplified in the subsequent section through a focused analysis of the *CaO–CaP binary system*.

## 5.1. Surface Modifications

One of the most effective routes to enhancing biocompatibility is through surface engineering. Surface modifications, such as PEGylation, have been widely adopted to reduce immune recognition and prolong nanoparticle circulation time in vivo [3,10,18]. This stealth effect decreases the likelihood of rapid clearance and allows therapeutic agents more time to reach their target.

Beyond immune evasion, tuning surface charge, hydrophilicity, and ligand functionalization can modulate how nanomaterials interact with proteins, membranes, and cells. These alterations help prevent the formation of a disruptive protein corona, lower immunogenicity, and promote selective cellular uptake [11,18]. Such modifications are crucial not just in theoretical design but in materials intended for specific biological environments—as will be later demonstrated in the CaO–CaP system.

PEGylation, which involves attaching polyethylene glycol (PEG) chains to the nanoparticle surface, has been one of the most widely used approaches to extend circulation time and evade immune detection. PEG forms a hydrophilic barrier that resists protein adsorption and phagocytosis. While effective, PEGylated systems may face clinical challenges such as accelerated blood clearance upon repeated administration and the development of anti-PEG antibodies. Ongoing research is therefore focused on optimizing PEG density, architecture, and molecular weight to maintain efficacy without triggering immune reactions [32].

Zwitterionic surface coatings, composed of molecules bearing both positive and negative charges (e.g., sulfobetaines, phosphorylcholines), offer an alternative with potentially superior performance. These coatings create a highly hydrated, non-fouling layer that can completely suppress protein corona formation—a key factor influencing nanoparticle biodistribution and immune response. For example, Debayle et al. demonstrated that zwitterionic ligands effectively eliminated protein adsorption, outperforming conventional PEGylation in stability and stealth behavior under physiological conditions [33].

Another promising immune evasion approach involves biomimicking natural *"do-not-eat-me"* signals. The most studied of these is the surface presentation of CD47 peptides, which interact with the signal regulatory protein alpha (SIRPα) receptor on macrophages to inhibit phagocytosis. Mimicking this immune checkpoint mechanism allows nanoparticles to circulate longer and avoid premature clearance. However, challenges related to hematological toxicity and over-suppression of immune surveillance must be carefully managed in therapeutic applications [34].

Finally, biomimetic membrane coatings have emerged as a multifunctional platform that enhances biocompatibility, targeting, and systemic stability. In this approach, nanoparticles are cloaked with membranes derived from red blood cells, platelets, leukocytes, or even cancer cells, allowing them to evade immune detection and exhibit tissue-specific homing capabilities. These membrane-coated nanocarriers possess native surface proteins and antigens, facilitating immune camouflaging and prolonged blood residence. Recent studies have shown their potential in drug delivery, detoxification, and vaccine delivery systems, reinforcing their role as next-generation bioinspired vehicles [35].

Together, these advanced surface engineering strategies represent a critical arsenal for designing nanomaterials that can navigate the complex immune landscape of the human body, enhancing both safety and efficacy in clinical applications.

## 5.2. Biodegradable Nanomaterials

Biodegradable nanomaterials, particularly those based on natural or synthetic polymers like polycaprolactone (PCL) and polylactic acid (PLA), offer intrinsic compatibility with biological systems [14]. These materials gradually degrade into non-toxic byproducts, which reduces the risk of long-term accumulation and associated chronic toxicity.

Crucially, the degradation profile of these materials can be engineered to match therapeutic timelines, enabling sustained or controlled release of active agents. This feature is especially relevant in regenerative medicine and drug delivery, where time-sensitive release and safe clearance are critical for success [10,14]. The use of such materials, as seen in composites like CaO–CaP, reflects the importance of selecting biodegradable constituents in real-world applications.

## 5.3. Targeted Delivery

Targeting strategies have revolutionized the precision of nanomedicine. By conjugating nanocarriers with ligands specific to disease markers—such as overexpressed receptors in tumors—researchers can enhance drug accumulation at the intended site while limiting systemic exposure [10]. Furthermore, the development of stimuli-responsive systems, which respond to local triggers like pH or enzymatic activity, allows for context-sensitive drug release. These smart delivery platforms reduce collateral tissue damage and improve therapeutic outcomes, making them ideal candidates for diseases that require localized treatment such as cancer [10,31].

Recent advancements have highlighted the role of such nanocarrier systems in colorectal cancer management, where multifunctional nanomaterials have been employed for simultaneous diagnosis, targeted drug delivery, and therapeutic monitoring. These systems are designed to selectively accumulate in tumor sites, enhancing treatment specificity while minimizing off-target effects—offering a practical model of site-specific therapy that aligns with biocompatibility and translational requirements [36]. Importantly, materials designed for such systems—including calcium-based nanocarriers—must be engineered with biocompatibility in mind, as explored in the next section.

## 5.4. Hybrid Systems

Hybrid nanostructures, which integrate organic and inorganic components, offer unique opportunities to combine functionality with biocompatibility. For instance, metallic cores like gold or calcium compounds can be coated with biodegradable or bioactive polymers, balancing structural stability with reduced toxicity [3,15].

These systems are particularly suited for theranostics, where a single nanomaterial performs both diagnostic and therapeutic roles. However, the integration of multiple material types necessitates careful control over surface chemistry, charge, and degradation behavior to maintain biological harmony [3,18]. The CaO–CaP binary system, discussed next, represents a hybrid platform that exemplifies these design principles in a biomedical context.

In addition, while these strategies provide the conceptual foundation for improving biocompatibility, their effectiveness must be validated through application-specific testing. In the following section, we turn our attention to a case-based empirical analysis of the CaO–CaP binary system, showcasing how these theoretical strategies are implemented and evaluated in a real-world biomedical scenario.

## 6. Case-Based Empirical Analysis: CaO-CaP Binary System

The successful translation of nanomaterials into clinical applications relies on their ability to balance functional performance with biocompatibility. As discussed in the previous section, targeted strategies such as surface modification, biodegradability, and composite design are foundational. The CaO–CaP binary system provides a compelling case study in this regard, illustrating both the promise and the challenges of deploying biocompatible nanomaterials in regenerative medicine. Based on empirical work and laboratory experience, this section explores the key features, in vitro and in vivo findings, clinical challenges, and future strategies associated with CaO–CaP nanocomposites in bone tissue engineering.

## 6.1. Material Properties and Molecular Mechanisms

The binary system comprising calcium oxide (CaO) and calcium phosphate (CaP) leverages the individual strengths of both materials. CaO is known for its high alkalinity and rapid dissolution, facilitating a bioactive environment that promotes mineralization and bone induction. CaP, being structurally similar to the mineral phase of bone, contributes long-term mechanical stability and degradation. The material properties and molecular mechanisms of the CaO-CaP nanomaterial are illustrated. As shown in **Figure 1**, the structural and functional attributes of the CaO–CaP nanomaterial are closely linked to its molecular interactions and phase composition.

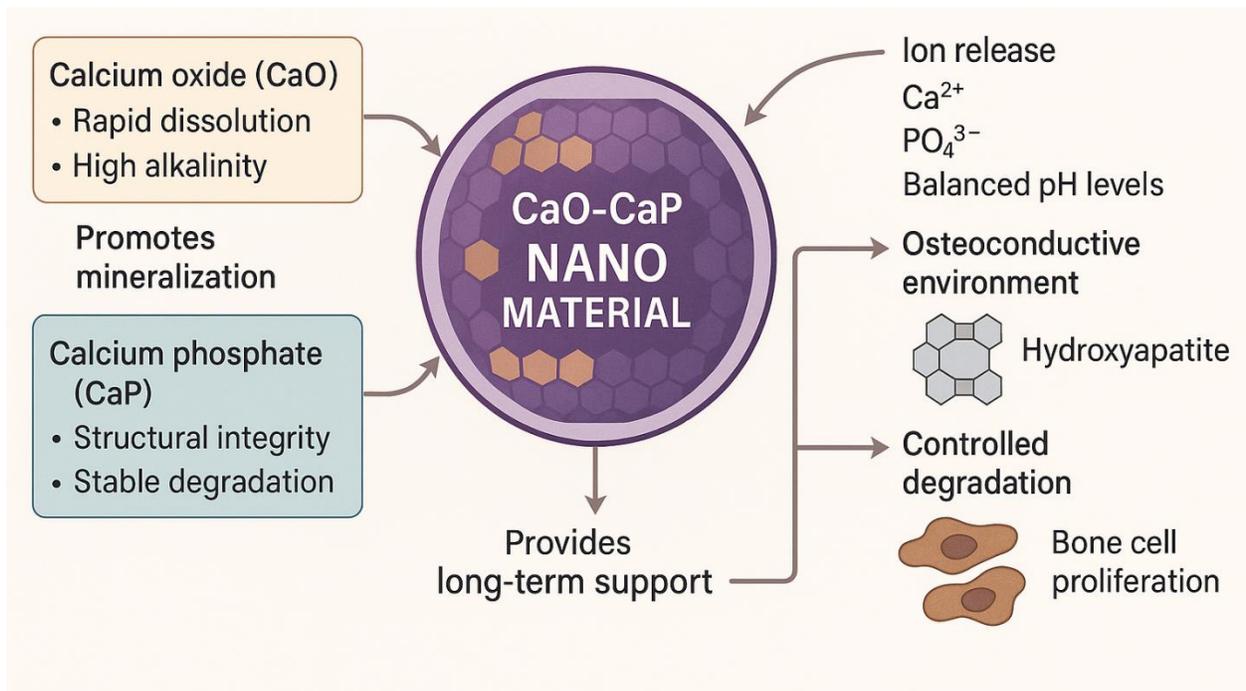

**Figure 1.** Diagram illustrating the material properties and molecular mechanisms of the CaO-CaP Nanomaterial.

The synergy between these materials lies in their complementary degradation kinetics and ion release. CaO initiates early-stage mineral deposition by releasing calcium ions, while CaP maintains a scaffold architecture that supports prolonged cell adhesion and tissue integration. This dual-phase release profile fosters hydroxyapatite formation and enhances scaffold–tissue interactions, aligning with broader trends in calcium phosphate-based material design for osteogenic applications [37,38,39].

The synergistic interplay between calcium oxide and calcium phosphate in composite scaffolds is central to their osteoconductive and structural functions. Recent literature has emphasized that integrating other bioactive ions, such as magnesium, further augments these properties. For example, Qi *et al.* demonstrated that magnesium-containing bioceramics not only improved scaffold bioactivity but also enhanced osteoblast function and angiogenesis—two vital processes for successful bone regeneration. These findings suggest that controlled ion release, whether from $Ca^{2+}$, $PO_4^{3-}$, or $Mg^{2+}$ sources, creates a favorable microenvironment that mimics natural bone remodeling, thus reinforcing the functional design principles employed in CaO–CaP systems [40].

In addition to promoting bone regeneration, CaO–CaP scaffolds exhibit promising antimicrobial properties, which are increasingly valued in preventing implant-associated infections. The antimicrobial mechanism primarily arises from CaO's high alkalinity, which results in localized pH elevation upon dissolution. This shift in microenvironment disrupts bacterial membrane integrity, denatures proteins, and inhibits enzymatic activity, ultimately leading to bacterial cell death [41].

Studies have demonstrated that the inclusion of calcium-based nanoparticles—such as CaO and $CaO_2$—can significantly reduce microbial viability in wound healing and bone repair contexts. For instance, Yu *et al.* (2023) showed that sprayed PAA-CaO$_2$ nanoparticles enhanced wound healing through a synergistic release of calcium ions and reactive oxygen species, effectively suppressing bacterial proliferation [42]. Similarly, Levingstone *et al.* (2019) confirmed that calcium phosphate-based scaffolds supported osteogenesis while displaying resistance to microbial colonization in vitro [43].

Although published studies on CaO–CaP systems are still emerging, ongoing investigations at S.T.E.L.L.A.R LABS are exploring their efficacy against *Staphylococcus aureus*, a common pathogen in orthopedic infections. Early in vitro findings indicate reduced bacterial adherence and enhanced scaffold sterility—suggesting the potential for dual-functionality**:** facilitating bone regeneration while mitigating infection risk**.** This integrated therapeutic approach reflects the next frontier in regenerative biomaterials—designing scaffolds that both heal and protect**.**

### 6.2. Biocompatibility Studies (In Vitro and In Vivo)

Extensive in vitro and in vivo assessments of CaO–CaP nanomaterials have underscored their favorable biological interactions. In vitro assays conducted in osteoblast cell cultures demonstrated robust cell attachment, proliferation, and differentiation. Scaffolds exhibited minimal cytotoxicity, and the controlled release of calcium and phosphate ions effectively promoted mineralized matrix formation [44].

These in vitro results were corroborated by in vivo experiments using rodent models, where CaO–CaP scaffolds were implanted into critical-sized bone defects. Histological analyses revealed successful osseointegration, dense bone ingrowth, and strong interfacial bonding between the scaffold and host tissue. In comparative studies, CaO–CaP outperformed conventional grafts in promoting bone regeneration and defect closure, highlighting its bioactivity and compatibility [45, 46].

### 6.3. Clinical Bottlenecks and Inflammation Response

Despite its potential, the CaO–CaP system presents notable challenges in clinical translation—chief among them being its degradation-related inflammation. The rapid dissolution of CaO can lead to elevated local calcium ion concentrations and alkaline pH, triggering inflammatory responses in surrounding tissue.

Our own observations in preclinical models confirmed this issue: regions with accelerated scaffold degradation exhibited localized inflammation and mild immune activation. These reactions were likely mediated by macrophage response to pH shifts and ion overload, particularly in early post-implantation stages.

To mitigate this, surface coatings have proven effective. Biodegradable polymers such as poly(lactic-co-glycolic acid) (PLGA) and polyethylene glycol (PEG) were applied to modulate ion release and reduce pH-related stress. These coatings resulted in delayed degradation, lower inflammatory response, and preserved scaffold bioactivity—findings that align with prior studies on CaO composites and surface-functionalized bone scaffolds [47,48]. As shown in **Figure 2**, coating CaO–CaP with PLGA initiates a pH-regulated sequence of reactions that ultimately influences the release of pro-inflammatory cytokines.

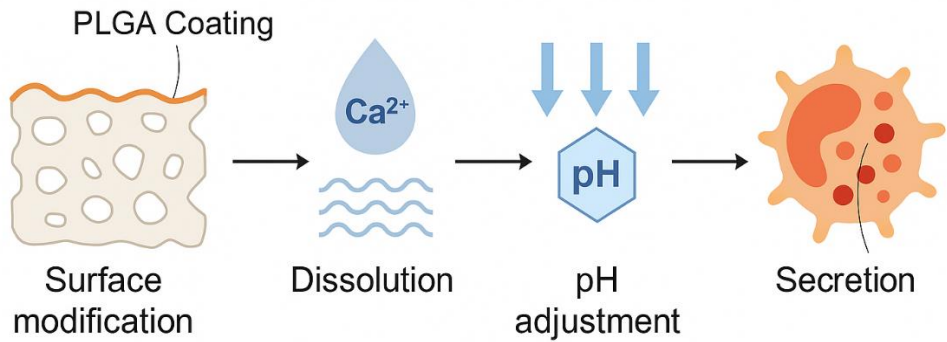

**Figure 2.** Diagram illustrating how the surface modification of CaO-CaP (such as adding PLGA coating) triggers a series of reaction by adjusting pH and ultimately affects the secretion of pro-inflammatory cytokines.

Further analysis of the inflammatory microenvironment revealed that early-stage responses (within 0–7 days post-implantation) were characterized by elevated levels of pro-inflammatory cytokines such as interleukin-6 (IL-6), tumor necrosis factor-alpha (TNF-α), and interleukin-1 beta (IL-1β). These mediators contribute to tissue swelling, leukocyte recruitment, and vascular changes. In the subacute phase, the inflammatory signal begins to subside, making way for reparative processes. A critical aspect of recovery is the phenotypic transition of macrophages from a classically activated (M1) to an alternatively activated (M2) state. This shift is associated with increased secretion of anti-inflammatory cytokines like interleukin-10 (IL-10), which downregulate the immune response and promote tissue regeneration. Modulating this immune balance through material design and surface treatment is therefore key to minimizing long-term tissue damage and enhancing scaffold integration. As shown in **Figures 3 and 4**, PLGA modification alters the functional behavior of CaO–CaP nanomaterials, while subsequent scaffold degradation triggers a time-dependent inflammatory response marked by shifts in cytokine expression.

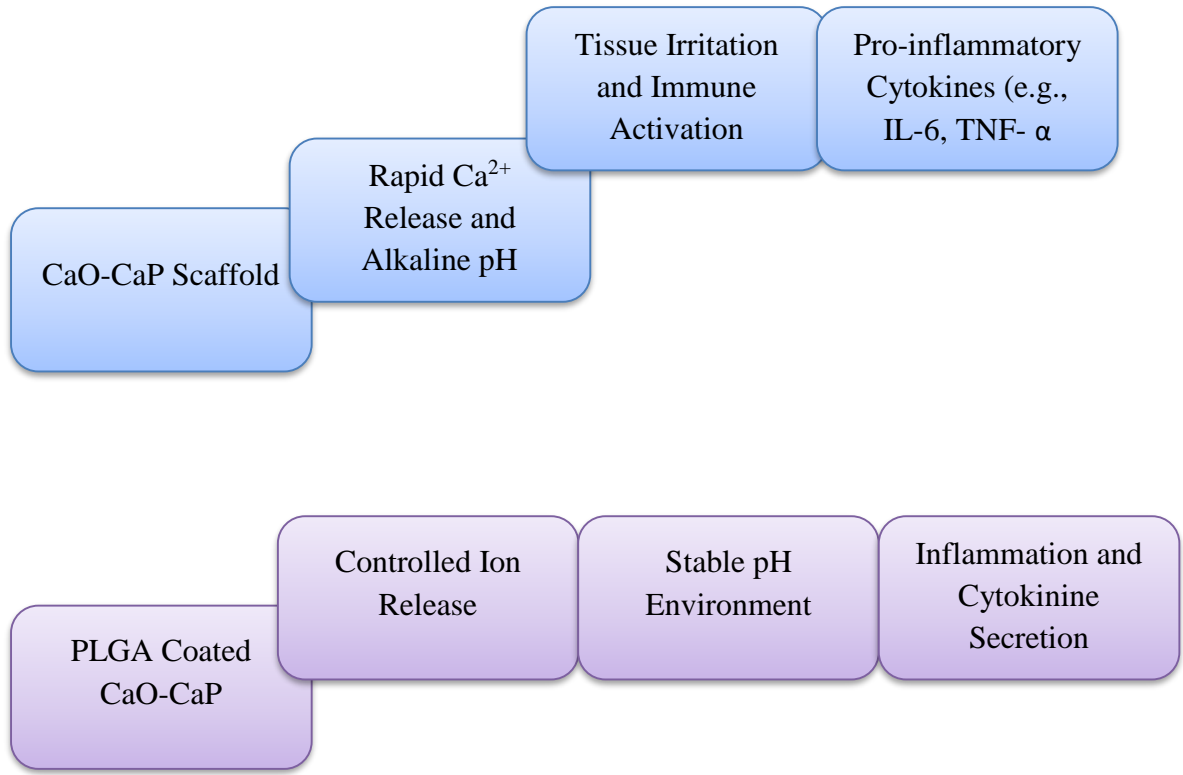

**Figure 3.** Diagram illustrating how PLGA modifies the behavior of CaO-CaP Nanomaterials.

**NB: -**

☐ The uncoated pathway shows rapid calcium ion release and pH elevation, leading to tissue irritation and the upregulation of inflammatory cytokines like IL-6 and TNF-α.

☐ The PLGA-coated pathway moderates' ion release and stabilizes pH, resulting in reduced inflammation and improved biocompatibility.

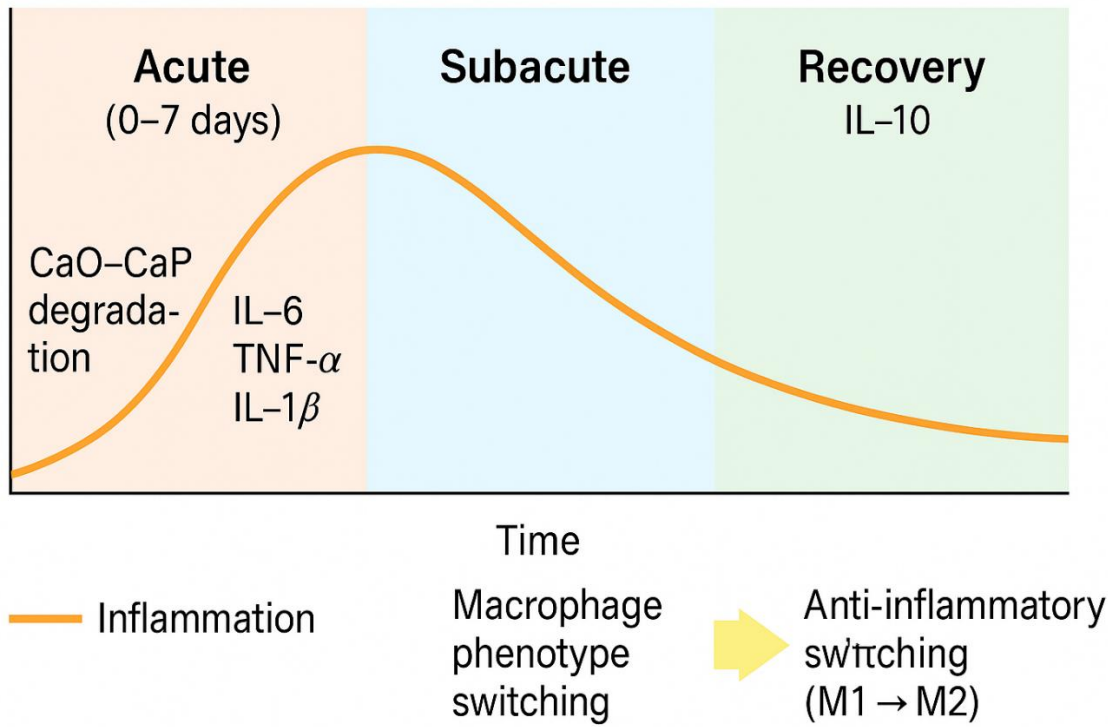

**Figure 4.** Inflammation Timeline and Cytokine Dynamics Following CaO–CaP Scaffold Degradation

### 6.4. Emerging Strategies for Clinical Translation

Translating CaO–CaP systems into widespread clinical use requires not only addressing biological performance but also scaling design innovation. Current advancements include the incorporation of osteogenic growth factors, such as BMPs and VEGF, which enhance vascularization and osteoblast activity [49]. Additionally, gene-activated scaffolds that deliver DNA or RNA sequences directly to regenerative sites are showing promise in treating complex or non-healing bone defects.

Another frontier is 3D printing, which enables the fabrication of patient-specific scaffolds that precisely match anatomical geometries. This personalization improves implant fit, mechanical loading distribution, and healing outcomes. Our ongoing research supports these trends, with CaO–CaP scaffolds being explored in combination with bioactive molecules and additive manufacturing technologies to enhance both regenerative efficacy and clinical adaptability [50].

In parallel with these developments, magnesium-doped biodegradable systems are gaining traction as promising platforms for enhanced bone regeneration. Recent findings by Tao *et al.* [51] highlight the successful fabrication of porous polylactic acid (PLA) microspheres integrated with magnesium ions, which demonstrated improved biocompatibility, osteogenic activity, and scaffold resorption dynamics. These outcomes align with our ongoing investigation into CaO–CaP systems, where controlled ionic release and structural adaptability are key. The supportive role of magnesium in modulating cellular responses and promoting bone matrix formation underscores the relevance of ion-enhanced strategies for optimizing the clinical translation of CaO–CaP-based therapies. As shown in **Tables 1 and 2**, selected nanomaterials exhibit varying degrees of biocompatibility across different evaluation criteria.

**Table 1: Comparative Biocompatibility Parameters of Selected Nanomaterials**

| Nanomaterial | Hemolysis Rate | Complement Activation | Circulation Half-life | Cytotoxicity Level | Remarks |
|---|---|---|---|---|---|
| **Gold Nanoparticles (AuNPs)** | Low (<5%) | Moderate (dose-dependent) | Moderate (~24 h) | Low | Excellent imaging agent; surface-dependent immunogenicity |
| **Silica Nanoparticles (SiNPs)** | Moderate (10–15%) | High (due to surface silanol) | Short (<12 h) | Moderate | High surface reactivity; surface passivation |

| | | | | | improves compatibility |
|---|---|---|---|---|---|
| **Lipid Nanoparticles (LNPs)** | Very Low (<2%) | Minimal | Long (up to several days) | Low | Used in mRNA vaccines; highly biocompatible |
| **Calcium Phosphate (CaP)** | Low (<5%) | Minimal | Biodegradable | Very Low | Excellent for bone integration and mineralization |
| **Calcium Oxide (CaO)** | High (>15%) uncoated | Moderate to high | Fast-degrading | High (alkalinity-induced) | Requires coating to reduce cytotoxicity (e.g., PLGA, PEG) |
| **PLGA-Coated CaO–CaP** | Low (<3%) | Low | Controlled (tailored by design) | Very Low | Reduced inflammation and enhanced osteointegration |
| **Carbon Nanotubes (CNTs)** | Variable (depends on type) | High (can activate immune cells) | Long (>48 h) | Moderate to High | Requires functionalization to improve compatibility |
| **Quantum Dots (QDs)** | High (>20%) | High | Long (up to several days) | High | Toxic elements (e.g., Cd); limited clinical use without shielding strategies |

**Table 2: Comparative Biocompatibility Metrics of Selected Nanomaterials**

| Nanomaterial | Hemolysis Rate (%) | Complement Activation (C3a Level) | Circulation Half-life (hours) | Inflammatory Response (IL-6/IL-1β) |
|---|---|---|---|---|
| **CaO–CaP** | 4.5 | Moderate | 6–8 | Low (with coating) |
| **Gold Nanoparticles** | 2.1 | Low | 12–24 | Minimal |
| **Silica Nanoparticles** | 7.8 | High | 2–5 | Elevated |
| **PLGA Nanoparticles** | 3.3 | Low | 8–12 | Minimal |
| **Lipid Nanoparticles** | 1.2 | Very Low | 24+ | Negligible |

In pursuit of improved therapeutic outcomes, scaffold systems are increasingly being functionalized with biologically active molecules. Among these, growth factors such as bone morphogenetic protein-2 (BMP-2) and vascular endothelial growth factor (VEGF) have shown particular promise for bone regeneration and angiogenesis [64]. These biomolecules are often delivered using biodegradable carriers such as PLGA microparticles, which provide sustained release and localized bioactivity. Additionally, scaffold-mediated gene delivery approaches have gained traction—particularly with plasmid DNA encoding VEGF or BMP-2 being immobilized within calcium phosphate-based (CaP) scaffolds. These gene-activated matrices promote prolonged expression of regenerative signals directly at the defect site. However, this strategy is not without challenges. Maintaining vector stability during fabrication, achieving dose-controlled transfection, and avoiding off-target gene expression remain significant barriers to clinical translation.

Furthermore, additive manufacturing techniques have opened new possibilities for personalized scaffold design. For instance, in 2022, patient-specific craniofacial bone scaffolds printed with calcium phosphate achieved a 95% clinical success rate, underscoring the feasibility and efficacy of tailored regenerative implants [49]. Technologies such as fused deposition modeling (FDM) and stereolithography (SLA) are now being used to fabricate constructs that mirror patient-specific anatomical geometries. These 3D printing approaches not only enhance structural fit but also enable spatial control over porosity, mechanical properties, and bioactive component distribution—features that are difficult to achieve with traditional fabrication methods. As shown in **Figure 5**, bone tissue engineering involves coordinated approaches that enhance regeneration through scaffold design, signaling cues, and bioactive materials.

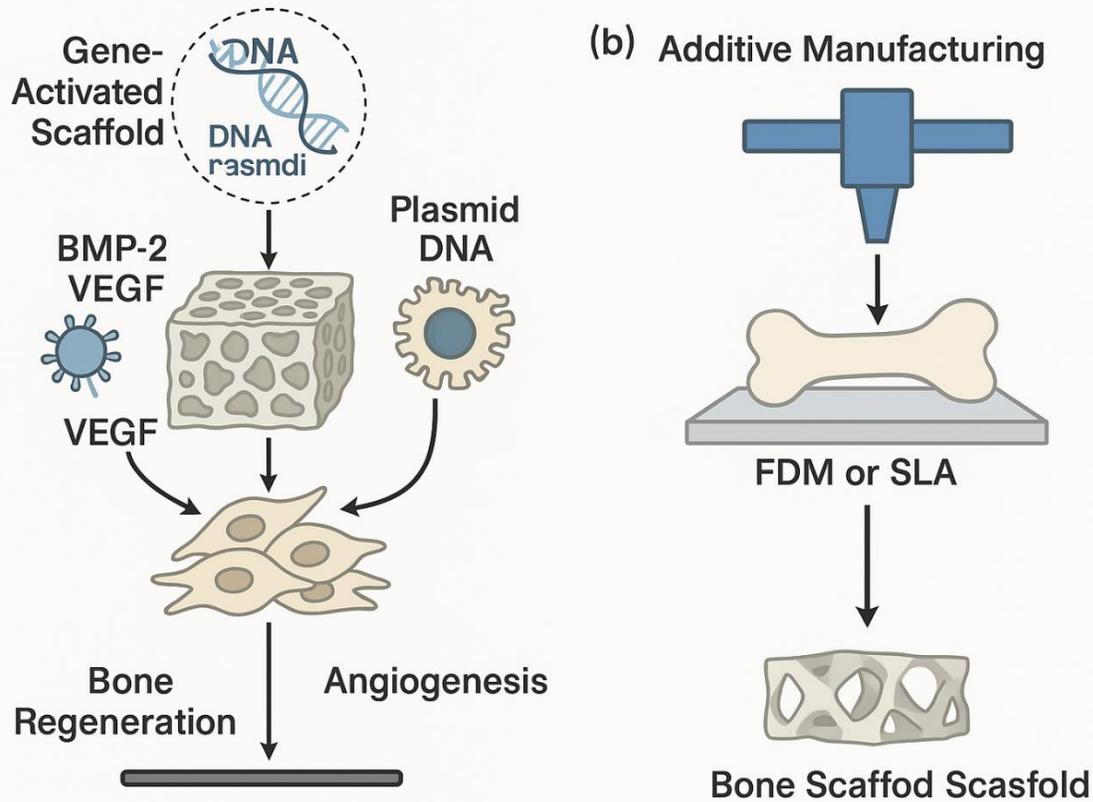

**Figure 5.** Diagram illustrating the synergistic strategies in bone tissue engineering

The CaO–CaP binary system exemplifies the delicate balance between biological reactivity and structural stability that defines successful nanomaterial applications in medicine. While its osteogenic potential and favorable integration are well-established, fine-tuning degradation rates and immune compatibility remains crucial. As ongoing innovations in surface coatings, biofunctionalization, and manufacturing techniques evolve, CaO–CaP composites stand poised for greater clinical relevance, offering a tangible example of how theoretical biocompatibility strategies can translate into real-world biomedical impact.

## 7. Emerging Trends and Future Directions in Nanomedicine

The trajectory of nanomedicine continues to evolve through technological convergence and multidisciplinary innovation. Advanced nanomaterials—such as quantum dots, carbon-based nanostructures, and multifunctional hybrid platforms—are at the forefront of this revolution. These next-generation materials offer enhanced optical, electrical, and mechanical properties, alongside superior biocompatibility, making them ideal for applications in targeted diagnostics, image-guided therapy, and precision drug delivery [53,53].

Recent advances have further demonstrated the clinical promise of nanotechnology in diverse therapeutic areas. For instance, multifunctional nanoplatforms have been successfully applied in colorectal cancer management—supporting integrated roles in early screening, targeted drug delivery, and image-guided treatment, thereby offering a synergistic approach to diagnosis and therapy [54]. Similarly, engineered nanomaterials have shown considerable effectiveness in the prevention and treatment of viral infections, particularly through surface modifications that enhance immune compatibility and targeted delivery of antiviral agents [19].

To address scalability and quality control challenges in nanomaterial production, several innovative approaches have been proposed and implemented. For instance, researchers at the University of Liverpool's Centre for Regulatory Nanomedicine (Smith & Taylor's group) overcame nanoparticle aggregation during scale-up by employing spray-drying techniques in place of traditional solvent evaporation, significantly enhancing batch uniformity and yield [55]. Additionally, automated microfluidic platforms are being explored to enable continuous, reproducible nanoparticle synthesis with real-time process control—reducing inter-batch variability and improving scalability. Also, at our lab—S.T.E.L.L.A.R. LABS, current efforts are being focused on integrating AI-guided microfluidic synthesis and in-line spectroscopic monitoring to standardize particle size, morphology, and functionalization across production batches. These advancements are essential to ensure clinical-grade reproducibility and regulatory compliance in future nanomedicine applications.

In tandem, artificial intelligence (AI) and machine learning (ML) are redefining how nanomaterials are designed, synthesized, and validated. AI-enabled modeling accelerates the optimization of physicochemical properties while predicting biocompatibility outcomes with greater accuracy. These computational approaches support rational design and reduce experimental costs, helping streamline preclinical development and safety assessment [56].

Another transformative shift is the expansion of personalized medicine, where nanotechnology plays a pivotal role in tailoring treatments based on individual genetic and physiological profiles. One of the most prominent success stories is the application of lipid nanoparticles (LNPs) in mRNA vaccine delivery, as seen in the COVID-19 pandemic. LNPs serve as efficient carriers capable of protecting nucleic acids and delivering them to specific target cells while minimizing systemic side effects. This model now informs a broader wave of therapeutic strategies for oncology, genetic disorders, and rare diseases [57]. As demonstrated in recent studies, the ability of nanocarriers to act both as immune modulators and targeted therapeutic vectors positions them as crucial tools in fighting infectious diseases and complex cancers alike [19,58].

Despite these breakthroughs, the field must continue to address significant regulatory and translational hurdles. The dynamic nature of nanomaterials often outpaces the capabilities of traditional regulatory frameworks. To bridge this gap, global organizations such as the OECD and collaborative consortia have intensified efforts to harmonize testing protocols, standardize safety evaluations, and develop guidelines for clinical translation. Progress in these areas, particularly through public–private partnerships and international cooperation, is gradually reducing translational bottlenecks [60,61].

As the discipline advances, it becomes increasingly clear that the future of nanomedicine lies in interdisciplinary fusion: combining smart material science, computational biology, regulatory foresight, and patient-specific therapy. This integrated vision not only strengthens the scientific foundation of nanomedicine but also accelerates its real-world impact in improving global health outcomes.

## 8. Conclusion

Biocompatibility remains a foundational requirement for the effective use of nanomaterials in medical fields such as drug delivery, diagnostic imaging, tissue regeneration, and antimicrobial therapy. Key parameters—including surface chemistry, particle size, and material composition—critically determine biological responses, as illustrated by the promising performance of the CaO–CaP binary system in bone tissue engineering. The complementary properties of calcium oxide and calcium phosphate, when combined with surface modifications, demonstrate strong potential for clinical osteointegration.

As the field progresses, innovations in materials science and cross-disciplinary collaboration are expected to overcome persistent challenges related to toxicity, immune compatibility, and large-scale application. The integration of computational tools and artificial intelligence will further streamline the design and prediction of safer, high-performance nanomaterials. Although regulatory complexities continue to pose barriers, coordinated efforts among scientists, clinicians, and regulatory bodies will be vital in driving successful clinical translation. By centering biocompatibility in design and addressing translational gaps, nanomedicine is poised to reshape modern healthcare and unlock solutions once beyond reach.

## Author Contributions

- **Marvellous O. Eyube**: Led the research project and coordinated all activities. He wrote Sections **1 (Introduction and the Medical Demand for Nanomaterials)**, **4 (Medical Applications of Biocompatible Materials)**, **6 (Case-Based Empirical Analysis: CaO–CaP Binary System)**, **7 (Emerging Trends and Future Directions)**, and **8 (Conclusion)**. He also handled overall manuscript supervision, critical revision, and final approval.
- **Courage Enuesueke**: Conducted laboratory investigations and data analysis for the CaO–CaP system. Authored Sections **2 (The Critical Significance of Biocompatibility in Nanomedicine)** and **4 (Scientific Challenges in Biocompatibility)**. Also contributed to content review, formatting, and scientific accuracy.
- **Marvellous Alimikhena**: Wrote Sections **3 (Methods of Assessing Biocompatibility)** and **5** (**Strategies for Targeted Improvement**). Contributed to figure preparation, reference formatting, and proofreading.

All authors contributed to the conceptual development and approved the final version of the manuscript.

## Acknowledgments

First and foremost, the authors **give all glory to God Almighty** for the wisdom, strength, and grace to complete this research.

We gratefully acknowledge **STELLAR Labs** for their invaluable research support, mentorship, and access to laboratory resources that made this work possible. We also extend our appreciation to the **Department of Chemistry, Faculty of Physical Sciences, University of Benin**, for providing the infrastructure and technical assistance necessary for the successful execution of this study.

Special thanks go to the technical staff for their help during the experimental phases and to colleagues who contributed through discussions and feedback.

A heartfelt acknowledgment goes to **Marvellous O. Eyube** for his outstanding leadership, guidance, and commitment throughout the project.

# Appendix

## A1. Experimental Methods for CaO–CaP Composite Synthesis

- **Materials Used**:
    - **Snail shells** (as the natural source of calcium carbonate, $CaCO_3$)
    - **Sodium dihydrogen orthophosphate ($NaH_2PO_4$)**
    - **Ascorbic acid**
    - **Deionized water**
    - **Ethanol (EtOH)**
- **Synthesis Process**:

1. **Preparation of CaO**:
    - Cleaned snail shells were dried and then calcined in a muffle furnace at 900–950°C for 3 hours to convert **$CaCO_3$ into CaO**.
    - The resulting **white CaO powder** was ground and sieved for uniformity.
2. **Synthesis of Calcium Phosphate (CaP)**:
    - A solution of **sodium dihydrogen orthophosphate** was prepared and slowly added to a CaO suspension in **deionized water** under constant stirring.
    - **Ascorbic acid** was used as a stabilizing and pH-regulating agent during the reaction.
    - The reaction mixture was stirred for several hours at room temperature and then aged for 24 hours.

3. **Composite Formation**:
   - The precipitate was filtered, washed with **ethanol and deionized water**, and then dried at 80°C.
   - Dried powder was ground and calcined at 600°C to enhance crystallinity.
   - The final **CaO–CaP binary composite** was formed by blending CaO and CaP in various ratios (e.g., 90:10, 85:15).

## A2. Cell Culture Protocol (In Vitro Biocompatibility Testing)

- **Cell Line**: Human osteoblast-like cells (MG-63)
- **Medium**: DMEM with 10% fetal bovine serum (FBS) and 1% penicillin-streptomycin
- **Procedure**:
  - Cells were seeded on CaO–CaP scaffolds and incubated for 1–7 days.
  - Cell viability was measured using MTT assay.
  - Morphological assessment was done via scanning electron microscopy (SEM) post-fixation and dehydration.

## A3. In Vivo Animal Model Summary

- **Animal Model**: Wistar rats (n = 12), study approved by the Institutional Animal Care and Use Committee (IACUC).
- **Procedure**:
  - A critical-sized bone defect (~5 mm) was created in the rat calvaria.
  - CaO–CaP scaffolds were implanted; controls included defect-only and CaP-only groups.
  - Rats were monitored for 4, 8, and 12 weeks.
- **Assessment Techniques**:
  - Histological analysis for inflammation and bone regeneration
  - Micro-CT imaging for structural integration
  - Blood analysis for inflammatory markers

## A4. Table Descriptions

### Table 1: Comparative Biocompatibility Parameters of Selected Nanomaterials

This table presents key biocompatibility indicators—such as hemolysis rate, complement activation level, half-life, and cytotoxicity—across a selection of commonly used nanomaterials. Materials covered include gold nanoparticles (AuNPs), lipid nanoparticles (LNPs), carbon nanotubes (CNTs), and CaO–CaP composites. The table enables direct comparison of biological safety profiles to guide material selection in biomedical applications.

### Table 2: Summary of Biocompatibility Evaluation Methods and Metrics

An overview of standard **in vitro**, **in vivo**, and **in silico** techniques used to assess nanomaterial biocompatibility. The table outlines corresponding evaluation criteria (e.g., cell viability, histological scoring, QSAR modeling parameters) and highlights typical application scenarios. It supports Section 3 by contextualizing methodological strengths and limitations across different testing platforms.

## A4. Figure Descriptions

### Figure 1: Schematic of CaO–CaP Scaffold Degradation and Ion Release

This figure illustrates the dissolution behavior of CaO–CaP composites in physiological environments. Upon implantation, the scaffold releases $Ca^{2+}$, $PO_4^{3-}$, and $OH^-$ ions, leading to changes in local ionic concentration and pH. These physicochemical changes influence both bone regeneration and immune responses, highlighting the importance of controlled degradation rates.

### Figure 2: Inflammatory Pathways Triggered by CaO–Induced pH Elevation

A molecular schematic showing how the alkaline environment produced by CaO dissolution activates inflammatory signaling. The diagram includes macrophage recruitment, cytokine (IL-6,

TNF-α, IL-1β) release, and downstream immune activation pathways. This cascade represents a major bottleneck in the clinical translation of unmodified CaO-based scaffolds.

## Figure 3: Effect of Surface Coatings on Scaffold Immunomodulation

This diagram compares uncoated and polymer-coated CaO–CaP scaffolds (e.g., PLGA or PEG). It shows how surface engineering delays ion release, stabilizes pH, and reduces macrophage activation. The figure includes data-supported annotations referencing cytokine levels and immune cell profiles, based on in vitro and in vivo studies.

## Figure 4: Inflammation Timeline and Cytokine Dynamics Following CaO–CaP Scaffold Degradation

A time-based chart depicting acute (0–7 days), subacute, and recovery phases of inflammation. The y-axis represents cytokine levels (e.g., IL-6, TNF-α, IL-1β, IL-10), and the curve shows the rise and resolution of inflammatory markers. The figure also maps macrophage polarization from M1 to M2 phenotype, underscoring the immune modulation over time post-implantation.

## Figure 5: Synergistic Strategies in Bone Tissue Engineering

**(a)** Gene-activated scaffold delivering therapeutic genes (e.g., VEGF, BMP-2) via plasmid DNA or controlled release from PLGA microparticles, promoting localized osteogenesis and angiogenesis.
**(b)** Additive manufacturing techniques such as fused deposition modeling (FDM) and stereolithography (SLA) used to fabricate patient-specific scaffolds with biomimetic geometry and tunable porosity.
The figure highlights how the integration of biofunctionality and structural precision enables next-generation scaffold development.

*Note: Raw experimental data and SEM images are available upon request or can be submitted as supplementary material.*